\documentclass{article}
\usepackage[margin=2.5cm]{geometry}
\usepackage{graphicx}
\usepackage{amsmath}
\usepackage{fixltx2e}
\usepackage{xspace}
\usepackage{subfigure}
\usepackage{setspace}
\usepackage{lineno}

\newcommand{\V}[2]{\ensuremath{\mathrm{V}_{#2}^{\mathrm{#1}}}}
\newcommand{\cc}{\ensuremath{\mathrm{cm}^3}\xspace}
\newcommand{\IPIP}{IPIP\xspace}

\title{\IPIP: A New Approach to Inverse Planning for HDR Brachytherapy by Directly Optimizing Dosimetric Indices}
\author{Timmy Siauw, Adam Cunha, Alper Atamt\"urk, I-Chow Hsu, Jean Pouliot, Ken Goldberg}

\begin{document}

\maketitle

\begin{abstract}
\noindent Purpose: Many planning methods for high dose rate (HDR) brachytherapy treatment planning require an iterative approach. A set of computational parameters are hypothesized that will give a dose plan that meets dosimetric criteria. A dose plan is computed using these parameters, and if any dosimetric criteria are not met, the process is iterated until a suitable dose plan is found. In this way, the dose distribution is controlled by abstract parameters. The purpose of this study is to improve HDR brachytherapy planning by developing a new approach that directly optimizes the dose distribution based on dosimetric criteria.

\noindent Method: We develop Inverse Planning by Integer Program (\IPIP), an optimization model for computing HDR brachytherapy dose plans and a fast heuristic for it. We used our heuristic to compute dose plans for 20 anonymized prostate cancer patient image data sets from our clinic database. Dosimetry was evaluated and compared to dosimetric criteria.

\noindent Results: Dose plans computed from \IPIP satisfied all given dosimetric criteria for the target and healthy tissue after a single iteration. The average target coverage was 95\%. The average computation time for \IPIP was 30.1 seconds on a Intel(R) Core$^{TM}$2 Duo CPU 1.67 GHz processor with 3 Gib RAM.

\noindent Conclusion: \IPIP is an HDR brachytherapy planning system that directly incorporates dosimetric criteria. We have demonstrated that \IPIP has clinically acceptable performance for the prostate cases and dosimetric criteria used in this study, both in terms of dosimetry and runtime. Further study is required to determine if \IPIP performs well for a more general group of patients and dosimetric criteria, including other cancer sites such as GYN.
\end{abstract}

\linenumbers

\section{Introduction}

High dose rate (HDR) brachytherapy is a radiation therapy for cancer. In HDR brachytherapy, radiation is delivered directly to the tumor site via a moving radioactive source in temporarily inserted catheters. Dose is controlled by altering the dwell times, the time spent at points along the catheters. Studies have shown that brachytherapy is a highly effective treatment \cite{Martinez, Yoshioka, Dinges, Galalae, Lachance, Mate, Blasko, Thomson}.
	
Inverse planning explicitly uses anatomical information and dwell positions when computing HDR brachytherapy dose plans. Clinical experience with HDR brachytherapy has led to organ doses that are correlated with biologically acceptable results \cite{RTOG}, more specifically, the eradication of the tumor with reasonable side effects to healthy tissue. Recently, these organ doses have become the dosimetric criteria that form the objective of many dose planning systems.

Many planning systems provide tools to quickly evaluate dose distributions and guide the user toward a final dose plan. However, finding a suitable dose plan may take several attempts. During dose planning, a set of computational parameters are hypothesized that will give a suitable dose plan (i.e. one that meets all dosimetric criteria). A dose plan is computed using these parameters, and if any dosimetric criteria are not met, the process is repeated until a suitable plan is found. In this way, the dose distribution is controlled by abstract parameters. The purpose of this study is to develop an inverse planning method for HDR brachytherapy that directly controls the dose distribution through dosimetric indices.

We formulate HDR dose planning as an optimization model known as a mixed integer program. A mixed integer program is an optimization model where some or all of the variables are restricted to taking on integer values. Dosimetric criteria can be directly incorporated into our model using integer programming constraints. Since the constraints in mixed integer programs are hard constraints, our model is guaranteed to meet the given dosimetric criteria if physically possible; it will be declared infeasible otherwise. However, many integer programs are difficult to solve in a reasonable amount of time \cite{Woolsey}. This is the case for our model, and in our initial tests, reasonably sized instances of our model did not provide an optimal solution, even given several hours.

HDR brachytherapy is an outpatient procedure. Catheters are inserted, a dose plan computed and delivered, and the catheters removed in the course of about one day. As a consequence, an HDR brachytherapy planning system must be able to compute a dose plan within several minutes to not interrupt clinical workflow. Since our model does not meet this runtime restriction, we develop a heuristic that does.

Although our heuristic converges to a solution quickly, it has three major limitations. First, as in all heuristic approaches, our heuristic is not guaranteed to compute an optimal solution for our model. Second, our heuristic is only guaranteed to comply with certain types of dosimetric criteria. Finally, within dosimetric criteria, our heuristic tends overdose healthy tissue, even for only small improvements to tumor dosing.

\begin{figure}[!ht]
  \begin{center}
    \includegraphics[width = 3.5in]{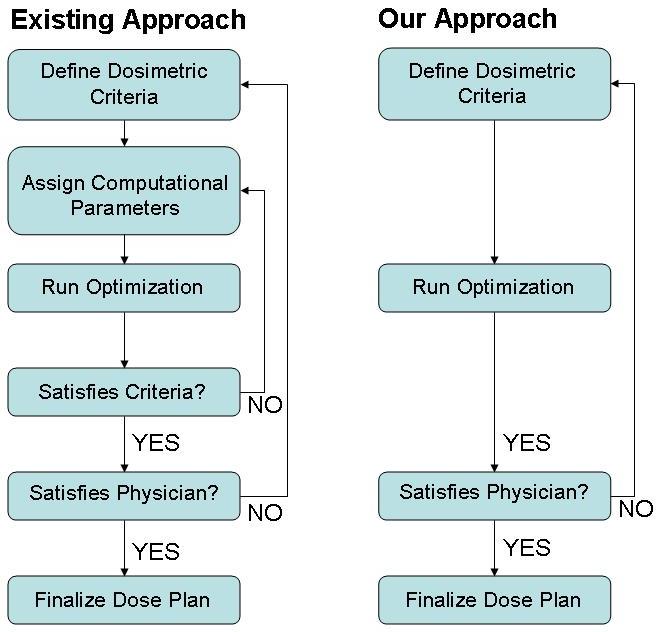}
  \end{center}
  \caption{Clinical workflow for current approach (left) and our approach (right). With current approaches, physicians iterate through dosimetric criteria and parameters. In our approach, iterations are only done using dosimetric criteria.}
  \label{workflow}
\end{figure}

Despite these limitations, we show that our heuristic approach can compute dose plans quickly, and for our patient cases, did not require any iterations beyond the first. Also, contrary to current approaches, if all dosimetric criteria are not met, physicians can iterate using different dosimetric criteria rather than computational parameters. A comparison of the workflow between the current approach and our approach is shown in Figure \ref{workflow}.

Although our model can be used for any brachytherapy cancer site where dosimetric criteria is known or desired, for this study we focus on the treatment of prostate cancer. In the following section, we give some background on inverse planning and integer programs.

\section{Background}

In HDR brachytherapy, the tumor site is referred to as the clinical target volume (CTV), and surrounding healthy organs are referred to as organs at risk (OAR). A physician prescribes a dose, Rx [Gy], that should be delivered to as much of the CTV as possible, subject to constraints on the OAR. The dose delivered to an entire organ is quantified by dosimetric indices. For example, the \V{Prostate}{100} is the prostate volume receiving at least 100\% Rx.

In inverse planning, organ volume is discretized into dose points. The dose received at the dose point is considered representative of the dose received in the near vicinity of the dose point. If dose points are evenly spaced, then a dosimetric index can be computed as the sum of all the dose points receiving over the dose threshold times the volume that a single dose point represents.

Clinical experience has determined ranges for dosimetric index values that are correlated with acceptable biological outcomes such as non-recurrence and minimal side effects to OAR. These dosimetric indices and their acceptable ranges constitute dosimetric criteria. For example, according to the RTOG-0321 dosimetric protocol \cite{RTOG}, when treating prostate cancer, the \V{Prostate}{100} (i.e. CTV coverage) should be greater than 90\% of the prostate volume and the \V{Rectum}{75} should be less than 1 \cc. The objective of HDR brachytherapy planning is to determine a set of source dwell times that achieves the given dosimetric criteria.

In iterative approaches to HDR brachytherapy planning such as in \cite{Lessard}, suitable dose plans are computed by assigning penalties to dose points and finding a dose plan that minimizes the total penalty. For any given dose plan, dose points are penalized for receiving dose in an undesirable range. For example, a dose point in the CTV could be penalized for receiving less than Rx, and a dose point in the rectum could be penalized for receiving more than 75\% Rx.

The importance of individual dosimetric criterion are represented by the relative magnitude of the penalty weights. For instance, if the penalty for the CTV is three times that of the rectum, then achieving CTV coverage is more important to the physician than controlling the dose to the rectum. Penalties are useful when a standard set of dosimetric criteria is not known because it allows physicians to balance CTV coverage versus OAR exposure in a single objective.

However, using penalties to achieve a prespecified set of dosimetric criteria can be difficult because penalties do not provide direct control over dosimetric indices. As a consequence, achieving dosimetric criteria with penalties becomes an iterative process. A set of penalties are hypothesized that will achieve dosimetric criteria; then a dose plan and dosimetry are computed. The process is repeated until a dose plan meeting all dosimetric criteria is achieved. Although a suitable dose plan can usually be found, it can take several iterations and is a time consuming process. Since standard dosimetric criteria are known for some cancer sites such as for prostate cancer, it is natural to utilize models that directly incorporate them.

The model and heuristic presented in this study utilize optimization models from mathematical programming. In the following paragraphs we give a brief overview of these models, and describe how they relate to brachytherapy dose planning.

A mixed integer program (MIP) is an optimization model of the form:

\begin{align*}
\mathrm{(MIP)}  & \mathrm{Maximize }\: c^T x\\
                & \mathrm{Subject\: to:}\\
                & A x \le b, \\
                & x_{i} \in R, i \in \{1,...,k\},\\
                & x_{i} \in Z, i \in \{k+1,...,n\},
\end{align*}

\noindent where the known parameters $c$, $A$, and $b$ are $n x 1$, $m x n$, and $m x 1$ matrices of real numbers, respectively, and x is an n x 1 vector of unknown variables. A feasible solution is an $x$ that satisfies every constraint. The term $c^T x$ is called the objective function. The goal of a mixed integer program solver is to find an optimal solution -- a feasible solution with the highest objective function value. If there are no variables in (MIP) that are restricted to be integers, then (MIP) reduces to a linear program (LP). LPs are efficiently solvable \cite{Bertsimas}, even on a personal computer.

Often, the mathematical structure of MIPs makes it difficult to find a provably optimal solution for them within a reasonable time \cite{Woolsey}. As a consequence, fast heuristic, or suboptimal, approaches are required to make usable applications.

Integer programs have been used to model prostate permanent-seed (PPI) brachytherapy and external beam dose planning \cite{Lu, Gallagher, Lee, Fox, Meyer, Ferris}. In particular, Ferris et al., 2002, used integer programs to constrain dosimetric indices. Our model extends these constraints to HDR brachytherapy which uses similar dosimetric indices. In many of these studies, extensive work was done to formulate customized algorithms to solve these models in a reasonable time. This study also presents a customized heuristic approach to compute fast but suboptimal solutions from our model. We present the model and heuristic in the following sections.

\section{Method and Materials}

\subsection{Model Formulation}

We develop a mixed integer program for HDR brachytherapy planning. A description of subscripts, parameters, and variables in this model are given for reference in Table \ref{Terms}.

\begin{table}
\centering
  \begin{tabular}{| r | l |}
    \hline
    Term          & Description                                          \\ \hline \hline
    $s$           & Subscript for organs.                                \\ \hline
    $i$           & Subscript for dose points.                           \\ \hline
    $j$           & Subscript for dwell positions.                       \\ \hline
    $G_s$         & The set of dose points in organ $s$.                 \\ \hline
    $P_{si}$      & The 3-D coordinates of dose point $i$ in $G_s$.      \\ \hline
    $N_s$         & The number of dose points in $G_s$.                  \\ \hline
    $T_j$         & The 3-D coordinates of dwell position $j$.           \\ \hline
    $N_T$         & The number of dwell positions for this patient.      \\ \hline
    $D_{sij}$     & The dose rate from $T_j$ to $P_{si}$.                \\ \hline
    $t_j$         & Dwell time at $T_j$.                                 \\ \hline
    $d_{si}$      & Dose at $P_{si}$.                                    \\ \hline
    $R_s$         & Threshold dose for $G_s$.                            \\ \hline
    $M_s$         & Maximum dose for $G_s$.                              \\ \hline
    $x_{si}$      & Indicator variable for $P_{si}$.                     \\ \hline
    $v_s$         & Dosimetric index for $G_s$.                          \\ \hline
    $L_s$         & Lower bound for $v_s$.                                \\ \hline
    $U_s$         & Upper bound for $v_s$.                                \\ \hline
  \end{tabular}
  \caption{\IPIP term definitions.}
  \label{Terms}
\end{table}

Dosimetric indices are estimated by sampling from a uniform grid of dose points that have been generated from evenly spaced anatomy image cross sections. The formulation of our model assumes the same set of dose points. The subscript "s" denotes organs, and "i" and "j" denote dose points and dwell positions, respectively. The set of dose points in an organ is $G_s$, and $P_{si}$ is the 3 dimensional coordinate of a dose point in $G_s$. The number of dose points in an organ is $N_s$. Dwell positions are denoted by $T_j$, and the number of dwell positions is $N_{T}$.

The dose rate parameters, $D_{sij}$ [cGy/s], are the dose received at $P_{si}$ for every second the source remains at $T_j$. $R_s$ [cGy] is the dose required for $P_{si}$ to be counted in the dosimetric index for $G_s$. The maximum allowed dose for dose points in $G_s$ is $M_s$ [cGy]. $L_s$ and $U_s$ are the upper and lower bounds for the dosimetric index for $G_s$, respectively.

The optimization variables are the dwell times, $t_j$ [seconds], the dose at each dose point $d_{si}$ [cGy], the dosimetric indices for each organ, $v_s$, and the indicator variables, $x_{si}$ [1]. The dwell times are continuous variables that represent the source time spent at $T_j$. The total dose received at $P_{si}$ is the sum of the contributions from every dwell position, $d_{si} = \sum_{j=1}^{N_T} D_{sij} t_j$. The indicator variables are binary variables that should have the following behavior:

\begin{equation*}
x_{si} = \left\{
    \begin{array}{cc}
            1 & \mathrm{if\ } d_{si} \ge R_s\\
            0 & \mathrm{otherwise}
    \end{array}
    \right.
\end{equation*}

\noindent Finally, the dosimetric index for $G_s$ is the sum of all the indicator variables, $v_s = \displaystyle\sum_{i=1}^{N_s} x_{si}$.

We dub the model Inverse Planning by Integer Program (\IPIP). It is presented in its entirety below.

\begin{align*}
&(\IPIP)                                                       &                       &     \\
&\mathrm{Maximize}\: v_0                                         &                       &     \\
&\mathrm{Subject\: to:}                                          &                       &     \\
&d_{si} = \displaystyle\sum_{j=1}^{N_{T}} D_{sij} t_j,         &\forall s,i\in G_s,    & (1) \\
&R_s x_{si}\le d_{si} \le R_s + (M_s - R_s) x_{si} - \epsilon, &\forall s,i\in G_s,    & (2) \\
&v_s = \displaystyle\sum_{i=1}^{N_s} x_{si},                   &\forall i\in G_s,      & (3) \\
&L_s \le I_s \le U_s                                           &\forall s,             & (4) \\
&t_j \ge 0                                                     &\forall j,             & (5) \\
&x_{si} \in \{0,1\}                                            &\forall s,i\in G_s.    & (6)
\end{align*}\\

The objective of \IPIP is to maximize $v_0$, CTV coverage. The first constraint (1) integrates the dose at $P_{si}$ from every dwell position. The purpose of (2) and (6) is to enforce the relationship between $d_{si}$ and $x_{si}$. For a given dose $d_{si}$ at $P_{si}$,  if $x_{si} = 1$, (2) reduces to $R_s \le d_{si} \le M_s - \epsilon$, and if $x_{si} = 0$, it reduces to $0 \le d_{si} \le R_s - \epsilon$. Conversely, if $d_{si} \ge R_s$, then $x_{si} = 1$, and if $d_{si} < R_s$, then $x_{si} = 0$. Therefore, $x_{si} = 1$ if and only if $d_{si} \ge R_s$, and the indicator variables behave as desired.

The parameter $\epsilon$ is a small number that is included so that the value of $x_{si}$ has no ambiguity when $d_{si} = R_s$. More specifically, if $d_{si}$ equals $R_s$, then 1 or 0 is valid for $x_{si}$ even though it should take the value 1 according to our definition. For very small $\epsilon$, (2) becomes an approximation to  $R_s x_{si}\le d_{si} < R_s + (M_s - R_s) x_{si}$, which ensures that $x_{si} = 1$ when $d_{si} = R_s$. If this level of precision is not required, then $\epsilon$ can be omitted from the formulation.

Constraint (3) integrates the dosimetric indices over all the indicator variables. Constraint (4) enforces dosimetric criteria. Since dose points lie on a uniform grid, each dose point is representative of an equal organ volume. Therefore, constraining the amount of organ volume receiving $R_s$ dose can be made equivalent to constraining the number dose points receiving $R_s$. For example if each dose point represents 0.1 \cc of organ volume, and no more than 1 \cc of organ can receive over the threshold dose, then the number of dose points in the organ that can receive over the threshold dose is 10 dose points.

Constraint (5) restricts the dwell times to be non-negative and (6) constrains the indicator variables to be binary.

\IPIP is a direct approach to finding dose plans with dosimetric criteria. However, our initial tests showed that it could not compute an optimal (or feasible solution) reliably for reasonably sized instances, even given several hours. Therefore in the following section, we develop a fast heuristic method for computing feasibly solutions.

\subsection{\IPIP Heuristic formulation}

The computational difficulty in solving \IPIP comes from the binary variables, $x_{si}$, that make \IPIP a MIP. We develop a heuristic to quickly determine values for the indicator variables in \IPIP. This is accomplished by first relaxing \IPIP using fewer and weaker constraints. The resulting optimization model is an LP, which can be solved quickly. The dose plan computed from this relaxation is analyzed, and based on this analysis, all $x_{si}$ from OAR (which have a dose upper limit) are set to 1 or 0. A doseplan is computed from the model with the set indicator variables. This dose plan is feasible for all upperbound dosimetric constraints (i.e. $v_s \le U_s$), but not necessarily feasible for lowerbound constraints (i.e. $L_s \le v_s$).

First, the binary restriction for the indicator variables in \IPIP is relaxed to allow all values between 0 and 1 (i.e. the constraint $x_{si} \in \{0,1\}$ becomes $0\le x_{si}\le 1$). Also constraints on dosimetric index values are removed. The resulting optimization problem is the following linear program referred to as the heuristic relaxation (HR).

\begin{align*}
\mathrm{(HR)}   & \mathrm{Maximize}\: \displaystyle\sum_{i=1}^{N_0} x_{0i}           &                   \\
                & \mathrm{Subject\: to:}\\
                & \displaystyle\sum_{j=1}^{N_{T}} D_{sij} t_j \ge R_s x_{0i},       &\forall i\in G_0,  \\
                & \displaystyle\sum_{j=1}^{N_{T}} D_{sij} t_j \le M_s - \epsilon,   &\forall s,i\in G_s,\\
                & t_j \ge 0                                                         &\forall j,         \\
                & 0 \le x_{0i} \le 1,                                               &\forall i\in G_0.
\end{align*}

Since dosimetric indices are not constrained, the indicator variables that make up these dosimetric indices are omitted in HR. The dose at any dose point is only limited by the maximum dose, $M_s$. The indicator variables for the CTV (i.e. $x_{0i}$) are retained and maximized in the objective.

In general, a dose plan computed from HR will not satisfy the dosimetric criteria for the OAR in \IPIP (i.e. $v_s > U_s$). In \IPIP, the value of the indicator variable $x_{si}$ determines if the dose at $P_{si}$ should be less than $R_s$ ($x_{si} = 0$) or $M_s$ ($x_{si} = 1$). Therefore, setting $x_{si}$ to 0 or 1 is equivalent to setting the dose upper limit for $P_{si}$ to $R_s$ or $M_s$ respectively. In the current solution to HR, all the OAR indicator variables have been set to 1, making the dose upper limit to these dose points $M_s$. The next step of this heuristic is to set all but $U_s$ of the OAR indicator variables to 0. Then, $v_s$ is guaranteed to be less than $U_s$ for these organs.

In this heuristic, the $U_s$ dose points in $G_s$ receiving the most dose retain a dose upper limit of $M_s$. The remaining dose points are restricted to receive less than $R_s$. This can be accomplished by adding the constraint $\displaystyle\sum_{j=1}^{N_{T}} D_{sij} t_j \le R_s - \epsilon$ to the $N_s - U_s$ dose points in $G_s$ that are receiving the least dose in the HR dose plan. The reason this allocation of indicator variables was chosen is as follows. The HR dose plan will have higher coverage than the \IPIP optimal solution because it has fewer constraints. We impose additional constraints on HR so that it will be feasible for \IPIP, and these additional constraints will reduce the coverage of the HR dose plan. Further restricting the coldest dose points will impact CTV coverage the least because they must be altered the least to be excluded from the dosimetric index for that organ. After these constraints are added, HR with the additional constraints is resolved as an LP.

A common dosimetric criterion is for CTV coverage to be more than a certain percentage of the CTV. This is represented in \IPIP by the constraint $L_0 \le v_0$. This constraint was removed in HR and is not enforced by the additional constraints added later on. As a consequence, meeting lower bound dosimetric index constraints is not guaranteed by this heuristic. However, usually the only dosimetric index with a lower bound constraint is for CTV coverage, and this dosimetric index is maximized the objective of HR. Therefore, if the upper bound constraints are not too stringent, this criterion is likely to be met or come close.

Our \IPIP heuristic is summarized as follows:\\
(1) Solve HR.\\
(2) For each $G_s$ except the CTV, let $P_{si*}$ denote the $N_s - U_s$ dose points receiving the least dose in $G_s$ in the HR dose plan solution.\\
(3) For every $P_{si}\in P_{si*}$ add the constraint $\displaystyle\sum_{j=1}^{N_{T}} D_{sij} t_j \le R_s - \epsilon$ to HR.\\
(4) resolve updated HR to get dose plan.

For the sake of brevity, we will not distinguish between \IPIP and this heuristic for the remainder of the paper.

\subsection{Patient Data Sets}
We applied \IPIP retrospectively to 20 prostate cancer patient cases. These patients were chosen to have a wide range of prostate volumes ranging from 23 to 103 \cc. For these patients, the physician implanted 14 to 18 catheters in the prostate with transrectal ultrasound (TRUS) guidance while the patient was under epidural anesthesia. Then Flexi-guide catheters (Best Industries, Inc., Flexi-needles, 283-25 (FL153-15NG)), which are 1.98-mm-diameter hollow plastic needles through which the radioactive source move, were inserted transperineally by following the tip of the catheter from the apex of the prostate to the base of the prostate using ultrasound and a stepper. A Foley catheter was inserted to help visualize the urethra.

After catheter implantation, a treatment planning pelvic CT scan was obtained for each patient. Three-millimeter-thick
CT slices were collected using a spiral CT. The CTV and OAR (urethra, rectum, and bladder) were contoured using the Nucletron Plato Version 14.2.6 (Nucletron B.V., Veenendaal, The Netherlands). The CTV included only the prostate and no margin was added. When segmenting the bladder and rectum, the outermost mucosa surface was contoured. The urethra was defined by the outer surface of the Foley catheter, and only the urethral volume within the CTV was contoured. The OAR were contoured on all CT slices containing the CTV and at least two additional slices above and below. Implanted catheters were also digitized.

For the contoured anatomical structures, dose points were generated by sampling from a uniform grid with 2 mm spacing in the x-y direction and 3 mm spacing in the z direction. Dose points were also generated in the body tissue (space between organs) with 4 mm spacing in the x-y direction. For the 20 cases, the total number of dose points for \IPIP ranged from 8860 to 25288. To reduce the computation required, every dose point more than 3.5 cm from the xy-centroid of the dwell positions were omitted in the optimization. However, they were included when computing the dosimetric indices. Our tests showed that removing these dose points had no effect on the dosimetric index values. The number of dose points after removing dose points ranged from 6144 to 15568.

\subsection{Dose rate calculations and clinical criteria}
The dose-rate contribution to a dose point from a source dwell position is a function of the distance between them. The dose-rate parameters were calculated as specified in the AAPM TG-43 dosimetry protocol \cite{TGb}. The radioactive material used in the source was $^{192}$Ir, and the prescription dose was 9.5 Gy.

The clinical criteria used in this study can be found in Table \ref{DoseReq}. The specifications for the \V{Prostate}{100}, \V{Urethra}{125}, \V{Rectum}{75}, and \V{Bladder}{75} are defined by RTOG-0321 \cite{RTOG}. RTOG specifies that the \V{Urethra}{125} be much less than 1\cc. We interpreted this to mean less than 0.1\cc.

The \V{Prostate}{150} is not explicitly constrained by the RTOG-0321 protocol. The \V{Prostate}{150} is restricted by the homogeneity index (HI) where:\\

\begin{equation*}
\mathrm{HI} = \displaystyle\frac{\V{Prostate}{100} - \V{Prostate}{150}}{\V{Prostate}{100}}.
\end{equation*}

\noindent It is generally preferred that $\mathrm{HI} \ge 0.6$; however, lower values of HI are acceptable if they allow for higher CTV coverage. For this study, we constrain HI to be greater than 50\% to maintain some control over the \V{Prostate}{150} while not being overly restrictive. Since we expect target coverage over 90\%, this restriction on HI can be enforced with $\V{Prostate}{150} \le 45\%$.

The restrictions that the \V{Urethra}{150}, \V{Rectum}{100}, \V{Bladder}{100}, and \V{Body}{200} be equal to 0 are not specified by RTOG but are considered preferable when possible at our clinic. The preference of the \V{Body}{200} comes from the desire to keep hot spots localized within the CTV.

These dosimetric index constraints represent our dosimetric criteria. They are summarized in Table \ref{DoseReq}.

\begin{table}
\centering
  \begin{tabular}{| c | l |}
    \hline
    Index & Requirement\\ \hline \hline
    \V{Prostate}{100}   & $\ge$ 90\%    \\ \hline
    \V{Prostate}{150}   & $\le$ 45\%    \\ \hline
    \V{Urethra}{120}    & $\le$ 0.1 \cc \\ \hline
    \V{Urethra}{150}    & = 0 \cc       \\ \hline
    \V{Rectum}{75}      & $\le$ 1 \cc   \\ \hline
    \V{Rectum}{100}     & 0 \cc         \\ \hline
    \V{Bladder}{75}     & $\le$ 1 \cc   \\ \hline
    \V{Bladder}{100}    & = 0 \cc       \\ \hline
    \V{Body}{200}       & = 0 \cc       \\ \hline
  \end{tabular}
  \caption{Dosimetric Criteria.}
  \label{DoseReq}
\end{table}

The parameters used for \IPIP to reflect these dosimetric criteria can be found in Table \ref{Params}. The values of 8 and 83 for $U_s$ represent the number of dose points in 0.1 \cc and 1.0 \cc, respectively, based on our grid spacing. The value of $U_s$ for the \V{Prostate}{150} is 45\% the number of dose points in the prostate. The dose to prostate dose points should be unrestricted so we have used an unrestrictively high number, 20000 cGy. We did this to avoid using infinity, which creates numerical problems with our optimization solver. The dosimetry in our results shows no dose points receiving this dose level for any patient.

\begin{table}
\centering
  \begin{tabular}{| c | c | c | c | c | c |}
    \hline
    Organ       & s & $R_s$ & $M_s$     & $U_s$      \\ \hline \hline
    Prostate    & 0 & 950   & 20000     & $N_0$      \\ \hline
    Prostate    & 1 & 1425  & 20000     & .45 $N_0$  \\ \hline
    Urethra     & 2 & 1140  & 1425      & 8          \\ \hline
    Rectum      & 3 & 712   & 950       & 83         \\ \hline
    Bladder     & 4 & 712   & 950       & 83         \\ \hline
    Body        & 5 & 1900  & 1900      & 0          \\ \hline
  \end{tabular}
  \caption{\IPIP Parameters}
  \label{Params}
\end{table}

\subsection{Method evaluation}
We used Matlab v. R2008b (Mathworks Inc.) to compute dose plans from \IPIP. The linear programming optimization was done using the Matlab interface for the Mosek Optimization Toolbox v.5, a medium to large scale optimization package. All computations were performed on a personal computer with Intel(R) Core$^{TM}$2 Duo CPU 1.67 GHz processor, 3 Gib RAM, and the Windows 32-bit operating system. We record the compliance of IP2H dose plans with our dosimetric criteria and the running time.

\section{Results}

\IPIP satisfied all our dosimetric requirements on the first iteration. For comparison, we also computed dose plans from Inverse Planning Simulated Annealing (IPSA), a clinically deployed dose planning system used at our clinic. All computational parameters for IPSA were the same as in Alterovitz, 2006 \cite{Alterovitz}. No computational parameters from IPSA were manipulated after the first iteration.

The compliance rate (out of 20 patients) for each individual dose objective is summarized in Table \ref{compliance}. Note that our constraints on dosimetric indices are hard constraints so even the slightest deviation from the requirement is considered a failure. The average CTV coverage from \IPIP was 95\%.

\begin{table}
\centering
  \begin{tabular}{| c | c | c | c |}
    \hline
                        & \multicolumn{2}{|c|}{Approach} \\ \hline
    Dosimetric Index    & IPSA      & \IPIP     \\ \hline \hline

    \V{Prostate}{100}   & 100\%     & 100\%     \\ \hline
    \V{Prostate}{150}   & 100\%     & 100\%     \\ \hline
    \V{Urethra}{125}    & 85\%      & 100\%     \\ \hline
    \V{Urethra}{150}    & 50\%      & 100\%     \\ \hline
    \V{Rectum}{75}      & 80\%      & 100\%     \\ \hline
    \V{Rectum}{100}     & 55\%      & 100\%     \\ \hline
    \V{Bladder}{75}     & 65\%      & 100\%     \\ \hline
    \V{Bladder}{100}    & 100\%     & 100\%     \\ \hline
    \V{Body}{200}       & 5\%       & 100\%     \\ \hline
    HI                  & 100\%     & 100\%     \\ \hline
    All Requirements    & 0\%       & 100\%     \\ \hline

  \end{tabular}
  \caption{IPSA and \IPIP compliance with dosimetric criteria. Percentage of patient dose plans (out of 20) that complied with each constraint for each approach. Note that the dosimetric constraints supplied were hard constraints, meaning even the slightest deviation from the limit is considered a failure.}
  \label{compliance}
\end{table}

On average, \IPIP computed a dose plan within 30.1 seconds. The maximum runtime was 86 seconds. Our \IPIP heuristic is made up of (1) solving a linear program, (2) sorting doses within each OAR, and (3) solving another linear program. Each one of these steps is polynomial time solvable, meaning that the number of computations grows slowly (i.e. as a polynomial) with respect to the size of the input. As a consequence, the runtime performance of our \IPIP heuristic is reliable for a more general audience of patients given that they have a similar number of dose points.

For comparison, IPSA computed a dose plan within 5 seconds on average, with a maximum runtime of 9 seconds. However, IPSA would have required further iterations to achieve dosimetric criteria. These additional iterations are not included in the running time results for IPSA.

We also conducted side tests to demonstrate cases in which \IPIP does not behave as desired. First we showed a set of dosimetric criteria that cannot be met with \IPIP by insisting that the \V{Rectum}{75}, and \V{Rectum}{75} be 0 \cc. Otherwise, dosimetric criteria were the same. The results for a single patient are shown in Table \ref{fail}. CTV coverage, which is lower bound constrained, is only 78\% which does not meet the 90\% requirement. However, all other criteria are fulfilled. In this case, additional iterations would be required using less stringent dosimetric criteria.

\begin{table}
\centering
  \begin{tabular}{| c | c |}
    \hline
    Dosimetric Index    & Value   \\ \hline \hline
    \V{Prostate}{100}   & 78\%    \\ \hline
    \V{Prostate}{150}   & 33\%    \\ \hline
    \V{Urethra}{125}    & 0.01\cc \\ \hline
    \V{Rectum}{75}      & 0 \cc   \\ \hline
    \V{Bladder}{75}     & 0 \cc   \\ \hline
  \end{tabular}
  \caption{Set of dosimetric criteria that cannot be met with \IPIP. The \V{Rectum}{75} and \V{Bladder}{75} were required to be 0. For this case, CTV coverage was only 78\% which does not meet our 90\% requirement. Less stringent dosimetric criteria must be given in the next iteration.}
  \label{fail}
\end{table}

Second, we showed a case where a dose plan computed from \IPIP complies with dosimetric criteria but may not be what the physician desires. We computed two dose plans from \IPIP using different sets of dosimetric criteria. The first dosimetric criteria set was the same that was used in our main experiment. The second dosimetric criteria set was more stringent with \V{Rectum}{75} and \V{Bladder}{75} less than 0.5 \cc. All other dosimetric criteria were the same as for the main experiment. The results are summarized in Table \ref{compare}.

\begin{table}
\centering
  \begin{tabular}{| c | c | c | c |}
    \hline
    Dosimetry           & Dosimetric Criteria 1         & Dosimetric Criteria 2 \\ \hline \hline
    \V{Prostate}{100}   & 96\%                          & 91\%                  \\ \hline
    \V{Prostate}{150}   & 33\%                          & 30\%                  \\ \hline
    \V{Urethra}{125}    & 0.01\cc                       & 0 \cc                 \\ \hline
    \V{Rectum}{75}      & 0.92 \cc                      & 0.45 \cc              \\ \hline
    \V{Bladder}{75}     & 0.95 \cc                      & 0.44 \cc              \\ \hline
  \end{tabular}
  \caption{Comparison of different dosimetric criteria for a single patient. For Dosimetric Criteria 1, dosimetric criteria was the same as in our main experiment. For Dosimetric Criteria 2, the \V{Rectum}{75} and \V{Bladder}{75} are restricted to be less than 0.5 \cc, which is more stringent. By restricting the \V{Rectum}{75} and \V{Bladder}{75}, CTV coverage drops by 5\% but with large reduction to OAR exposure.}
  \label{compare}
\end{table}

CTV coverage for the first dosimetric criteria is 96\%. To achieve this coverage, almost all of the dose allowance to the rectum and bladder are utilized (i.e. \V{Rectum}{75} and \V{Bladder}{75} is close to 1 \cc). With the second dosimetric criteria, CTV coverage drops to 91\%, but the overdosed volume to the rectum and bladder is cut in half.

Since the objective of \IPIP is to maximize coverage within limits to OAR dose, \IPIP will tend to utilize all of the OAR dose available to it. More specifically, \IPIP will make large increases to OAR dose, within its limits, even to make only small improvements to CTV coverage. In this case, the drop in CTV coverage (5\%) between the two dose plans is worth considering, but a physician may still prefer the second dose plan. However, determining the second dose plan cannot be accomplished without iterating through alternative dosimetric criteria.

\section{Conclusion}
\IPIP is a new approach that is more intuitive for physicians and less reliant on manual fine tuning. \IPIP allows physicians to directly specify desired dosimetric indices, rather than system-specific computational parameters. Our results demonstrate that \IPIP quickly generates dosimetry plans that are consistent with RTOG-0321 standard dosimetric criteria. Further study is required to determine if \IPIP can produce similar performance for a more general group of patients and dosimetric criteria, including other cancer sites such as GYN.

\section{Acknowledgements}

We would like to acknowledge Professor Laurant El Ghaoui and Sarah Drewes for their input during this work, and Justin Woo for conducting the initial tests to determine if \IPIP could be solved quickly. We would also like to acknowledge Professor Ron Alterovitz for his introduction into brachytherapy planning, and Professor Kris Hauser for his supervision during the formulation of \IPIP.

\end{document}